\documentstyle[aps,rotate,multicol,epsf]{revtex}
\def\beginwide{
        \end{multicols} \vspace*{-0.5cm} \noindent
        \rule{3.5in}{.1mm}\rule{.1mm}{5mm} \widetext \medskip }
\def\beginwidetop{
        \end{multicols} \vspace*{-0.5cm} \noindent
        \widetext \medskip }
\def\endwide{
        \hspace*{3.35in}~\rule[-5mm]{.1mm}{5mm}\rule{3.5in}{.1mm}
        \begin{multicols}{2} \vspace*{-1.0cm} \noindent }
\def\endwidebottom{
        \begin{multicols}{2} \vspace*{-1.0cm} \noindent }

\draft

\begin{document}
% ------------------------------------------------------------------------------
\title{Fixed points, stability and intermittency 
in a shell model for\\ advection of passive scalars
} 
\author{Julien Kockelkoren\footnote{Electronic Address: kockel@spec.saclay.cea.fr}$^1$
and Mogens H. Jensen\footnote[5]{Electronic Address: mhjensen@nbi.dk}}
\address{Niels Bohr Institute 
and Center for Chaos and Turbulence Studies, 
Blegdamsvej 17, DK-2100 Copenhagen {\O}, Denmark }
\date{18 November 1999}
\maketitle
% ------------------------------------------------------------------------------
\begin{abstract}
We investigate the fixed points of a shell model for the turbulent advection of passive
scalars introduced in \cite{jpv92}. 
The passive scalar field is driven
by the velocity field of the popular GOY shell model. 
The scaling behavior of the static solutions is found to
differ significantly from Obukhov-Corrsin scaling
$\theta_n \sim k_n^{-1/3}$ which is only recovered in the limit where 
the diffusivity vanishes, $D \rightarrow 0$. From the eigenvalue spectrum
we show that any perturbation in the scalar
will always damp out, i.e. the eigenvalues of the scalar are negative and
are decoupled from the eigenvalues of the velocity.
Furthermore we estimate Lyapunov exponents and the
intermittency parameters using a definition
proposed by Benzi et al.\ \cite{bppv85}. The full model is
as chaotic as the GOY model, measured by the maximal Lyapunov exponent,
but is more intermittent.

\end{abstract}
% ------------------------------------------------------------------------------
\pacs{PACS numbers: }

\begin{multicols}{1}
\smallskip
\smallskip

\section{Introduction}

In recent years much attention has been paid to the origin of
intermittency in fully developed turbulence. Progress has been made in
the somewhat simpler problem of the anomalous scaling in the passive 
advection of
a scalar quantity (e.g.\ temperature or the density of a pollutant).
Fundamental analytical results have been
obtained when the advecting velocity field was assumed to be Gaussian
and delta-correlated in time, the so-called Kraichnan model \cite{kra,Kupi,Falk}, 
and in the context of shell models \cite{wirth}. 
Here we consider the perhaps more realistic situation where the
passive
scalar shell model is driven by a non-Gaussian velocity field with finite
correlation time, namely the velocity field of the GOY model
\cite{Gledzer,OY}. 
In this case the model can be analyzed using standard techniques of
(low) dimensional dynamical systems, e.g.\ the study of bifurcations,
eigenvalue spectra and Lyapunov exponents.

In absence of convective effects the passive scalar field $\Theta$ 
is governed by the equation: 
\begin{equation}
\partial_t \Theta + (\vec{v}\cdot \vec{\nabla}) \Theta =
D \nabla^2 \Theta + F_{\Theta} \label{diffadv}
\end{equation}
Here $v$ is the velocity field, $F_{\Theta}$ is an external forcing 
and $D$ is the diffusion coefficient.

According to the analogue of the K41 theory \cite{k41} for the passive
scalar, developed by Obukhov and Corrsin \cite{obucor},
the structure functions 
\begin{equation}
S_p(\vec{r})=\langle |\Theta(\vec{x}+\vec{r}) - \Theta(\vec{x})|^p \rangle
\sim l^{H_p}
\end{equation}
(where $l=|\vec{r}|$) scale linearly with p, more precisely
$H(p)=p/3$.
Experimentally, however, one observes for $p>3$ \cite{baudet} strong
deviations from this. This is usually referred to as anomalous
scaling or intermittency.
The deviations seem to be even more pronounced than for the
structure functions of the velocity: the passive scalar is said to be
more intermittent.  

This paper is organized as follows: after having introduced the model,
we examine in section III the scaling of its fixed points and in section IV
their stability. These studies have already been performed for the GOY model 
\cite{kada1,Lohse,kada2}, 
to which the passive scalar model is coupled. We will review these
results for the sake of completeness. 
In section V we study the full dynamics of the model and investigate its 
chaotic and intermittent behavior.
 
\section{Shell model for passive scalars}
Shell models appear to
capture many properties of fully developed turbulent flows but are easier to study
than the Navier-Stokes equations (see \cite{mogens} for a review).
In this letter we study the passive scalar shell model proposed
in Ref. \cite{jpv92}. 
The multiscaling of the model is in good agreement
with experimental data \cite{baudet}. 
The GOY model has been studied intensively
\cite{Gledzer,OY,kada1,Lohse,kada2,mogens,JPV,benzi,pisarenko,bif,Gilson,Fridolin,jk,cris93}; 
the passive scalar model has attracted somewhat less
attention \cite{cris93}, \cite{wirth}, \cite{pmg}.

Both models are constructed in Fourier space, retaining only the complex modes 
$u_n$ and $\theta _n$ as a representative of all modes in the shell of wave 
number $k$ between $k_n = k_0 \lambda ^n$ and $k_{n+1}$.
One uses the following assumptions:
(i) the dissipation resp.\ diffusion is represented by a linear term of
the form: $-\nu k_n^2u_n$ resp.\ $-Dk_n^2\theta_n$ (ii) the nonlinear
terms of the form $k_n u_{n'} u_{n''}$ resp.\
$k_n \theta_{n'} u_{n''}$ with (iii) $n'$ and $n''$
among the nearest
and next-nearest neighbors of $n$ and (iv) in absence of forcing and
damping conservation of volume in phase space and conservation of
$\sum_n |u_n|^2$ and $\sum_n |\theta_n|^2$.
Moreover the scaling laws $u_n \sim k_n^{-1/3}$ and $\theta _n
\sim k_n^{-1/3}$ form a fixed point of the inviscid unforced equations.
 
The resulting equations are \cite{jpv92}:
\begin{eqnarray}
(\frac{d}{dt} + \nu k_n^2) \ u_n \ & =  & i \, (a_n k_n \,
u^*_{n+1}u^*_{n+2}
\, + \, b_n k_{n-1} u^*_{n-1}u^*_{n+1}  \nonumber \\
& & \, + \, c_n k_{n-2} u^*_{n-1}u^*_{n-2}) + f \delta_{n,4} \label{goy} \\
(\frac{d}{dt} + Dk_n^2) \ \theta_n \ & = & i \, (e_n(u^*_{n-1}\theta^*_{n+1} -
u^*_{n+1}\theta^*_{n-1}) \nonumber \\
& & \, + \, g_n(u^*_{n-2}\theta^*_{n-1} + u^*_{n-1}\theta^*_{n-2}) \, + \, 
\nonumber \\
& & h_n (u^*_{n+1} \theta^*_{n+2} +
u^*_{n+2}\theta^*_{n+1})) \ + \ \bar{f}\delta_{n,4} \label{passhell}
\end{eqnarray}
  
A possible choice for the coefficients is:
\begin{eqnarray}
a_n = 1 \quad b_n = -\delta \quad c_n = -(1 - \delta) \\ 
e_n = \frac{k_n}{2} \quad g_n = -\frac{k_{n-1}}{2} \quad
h_n=\frac{k_{n+1}}{2}
\end{eqnarray}
The boundary conditions are:
\begin{eqnarray}
b_1=b_N=c_1=c_2=a_{N-1}=a_N=0\\
e_1=e_N=g_1=g_2=h_{N-1}=h_N=0
\end{eqnarray}

For the parameters, we choose for example the following standard values:
\begin{eqnarray}
N=19, \lambda = 2, k_0 = \lambda ^{-4}, \nu = 10^{-6}, \nonumber \\
f=\bar{f}=5 \cdot 10^{-3} \cdot (1+i),  
D = 10^{-6}.
\label{standpar}
\end{eqnarray}
The free parameter $\delta$ is related to a second 
quadratic invariant which for
the canonical value $\delta=\frac12$ is similar to the helicity
\cite{kada1}.

These equations determine the evolution of the vector 
$(U,\Theta)=(\Re u_1, \Im u_1, \ldots, \Re u_N, \Im u_N, \Re \theta_1,
\ldots, \Im \theta_N)$ and thus form a $4N$
dynamical system.

\section{Scaling of fixed points}
A first step towards a full understanding of the model consists of an 
investigation of its static properties. In this section we examine the scaling
properties of the fixed points of 
equations (\ref{goy})-(\ref{passhell}). The major problem is to find the 
static solutions of
(\ref{goy}), those of (\ref{passhell}) can then be found easily because
$\dot{\theta}$ is linear in both $u$ and $\theta$. 

It was found in \cite{bif} that (\ref{goy}) in the unforced inviscid limit
allows three self-similar static solutions: the trivial fixed point
$u_n=0$, a ``Kolmogorov'' fixed point $u_n=k_n^{-1/3} g_1(n)$ and a
``flux-less'' fixed point $u_n=k_n^{-z} g_2(n)$, with $g_1(n)$ and
$g_2(n)$ any function of period three in $n$ and $z=(-\ln
_{\lambda}(\delta - 1) + 1)/3$. The corresponding fixed points of
(\ref{passhell}) are: $\theta_n = 0$,  $\theta_n = k_n^{-1/3} g_1(n)$  and
$\theta_n = k_n^{-\frac12 (1-z)}$.
We will here focus on the Kolmogorov fixed point which is believed to
be the most important for the dynamics \cite{bif} although it was
suggested that also the trivial fixed point might play a major role
\cite{Fridolin}.

We note that the static solution for $u_n=u_n e^{i \phi _n}$ 
can be turned into real form 
by a change of phase. Following
Sch\"orghofer et al.\cite{kada2} we choose the phases:
\begin{equation} \label{phases}
\phi_n= \left\{ \begin{array}{lll}
\frac14 \pi & {\rm for} & n=1,4,7,\dots\\
\frac18 \pi & {\rm for} & n=2,5,8,\dots\\
\frac98 \pi & {\rm for} & n=3,6,9,\dots\\
\end{array} \right.
\end{equation}
It can be shown that the static solution of $\theta$ picks up the
same phase as that of $u$.

As has been observed \cite{bif}, the dynamics of 
the system converges to the Kolmogorov fixed
point for $\delta < 0.379634$. 
When increasing $\delta$ the system undergoes a series of Hopf
bifurcations and becomes chaotic at $\delta = 0.38704$ \cite{bif,kada2,jk}.
In order to find the Kolmogorov fixed point one can thus vary
$\delta$ in small steps and refine the solution with Newton's method
\cite{Lohse}.

To study the scaling behavior of the static
solutions, we apply a much larger number of shells, using
the same parameter values as Kadanoff et al. \cite{kada2}:
\begin{equation}
N=90, \lambda = 2, k_0 = \lambda ^{-1}, \nu=D=10^{-3}16^{-26}, f=\bar{f}=1
\end{equation}
The forcing acts in this case on the first shell. 

\begin{figure}[htb]
\narrowtext
\unitlength= 0.0022\textwidth
\begin{center}
\begin{picture}(200,200)(0,0)
\put(0,0){\makebox(200,200){\epsfxsize=180\unitlength\epsffile{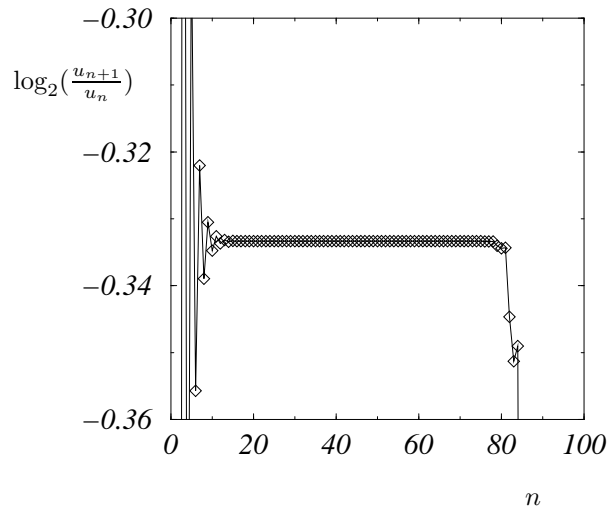}}}
\put(10,150){\makebox(0,0){$\log_2(\frac{u_{n+1}}{u_n})$}}
\put(165,10){\makebox(0,0){$n$}}
\end{picture}
\end{center}
\caption{Static solution of the {\sc goy} model at $\delta=0.5$ for
the parameters (11).
To improve the scaling behavior,
$\log_2(u_{n+1}/u_n)$ is averaged over three consecutive shells.
\label{ukad}}
\end{figure}

For the static solution of $u_n$
we obtain the same result as \cite{kada2}, see
figure \ref{ukad}, where $\log_2(u_{n+1}/u_n)$ is plotted  versus $n$.
It is averaged over a period three to get rid of the well known
period three oscillations.  
The solution is seen to follow Kolmogorov scaling. 

\begin{figure}
\narrowtext
\unitlength= 0.0022\textwidth
\begin{center}
\begin{picture}(250,250)(0,0)
\put(0,0){\makebox(200,200){\epsfxsize=180\unitlength\epsffile{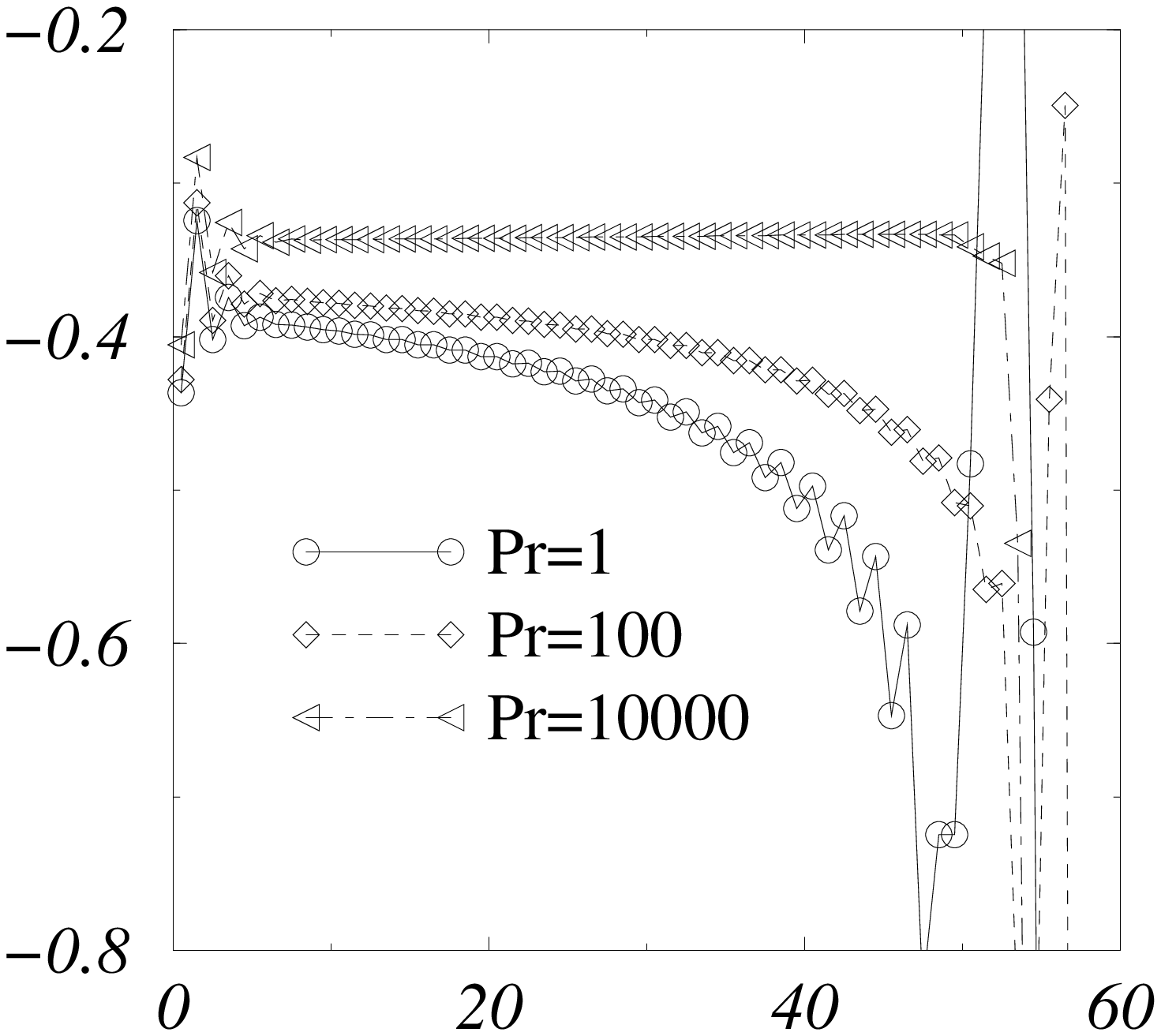}}}
\put(10,150){\makebox(0,0){$\log_2(\frac{\theta_{n+1}}{\theta_n})$}}
\put(160,10){\makebox(0,0){$n$}}
\end{picture}
\end{center}
\caption{Static solution, fixed point, of the passive scalar model
where 
$\log_2(\theta_{n+1}/\theta_n)$ (averaged  three consecutive shells) is plotted
versus the shell index $n$. Here, the number of shells is $N$=61.
\label{thstat}}
\end{figure}

Surprisingly however, the fixed point for $\theta_n$ (with the fixed point
of $u_n$ inserted) deviates strongly
from the Obukhov scaling for a finite value of the diffusivity $D$, 
as can be seen in figure \ref{thstat}. 
There is clearly not a well-defined power law scaling. 
This solution also displays the period three   
behavior. In the diffusive range the solution looks somewhat noisy,
presumably due to the boundary conditions. 
It is clear, that there is a ``slow'' bending in the diffusive regime
but as $D$ approaches zero, the curve becomes more and more flat 
and we recover the
Obukhov scaling in the limit $D \rightarrow 0$. 
We thus note that in contrast to the velocity-case where the viscosity only 
affects the
viscous range, for the passive scalar the diffusion seems to act on
the whole inertial range, at least for Prandtl numbers $Pr = \frac{\nu}{D}
\sim 1$. 
This might have its origin in the linear character of the problem.
%As a comparison, we can consider the dynamics of the full model at high
%values of the Prandtl number. Numerically, it is difficult to integrate
%for $N = 90$ but we study the case $N=27$ and $Pr = 10000$. The corresponding
%spectrum is shown in Fig. XX. Clearly, for this value of the Prandtl
%number, the spectrum bends over from a Kolmogorov scaling to a
%Bachelor regime $< \theta_n > \sim constant$ \cite{Olesen}. We therefore
%conclude, that the static behavior is very different from the dynamical
%behavior as a function of Prandtl number.

%This deviation is quite amazing since it seems to be absent in the dynamics
%of the model \cite{jpv92}. Experimentally one has observed a deviation
%from the Obukhov scaling, but less pronounced: $H(1)=0.37$ \cite{baudet}. 
%We note that if one increases the Prandtl number (i.e.\ decreases $D$,
%keeping $\nu$ constant), one recovers the
%Obukhov scaling. Thus the deviation has its origin in a finite value of the
%diffusion and hence 
%contradicts the widely accepted idea that dissipation
%range quantities like the molecular diffusion do not influence the
%inertial range behavior \cite{com}. 

%\begin{figure}[htb]
%\narrowtext
%\unitlength= 0.0022\textwidth
%\begin{center}
%\begin{picture}(200,200)(0,0)
%\put(0,0){\makebox(200,200){\epsfxsize=180\unitlength\epsffile{smt.eps}}}
%\end{picture}
%\end{center}
%\caption{Dynamics of the passive scalar+GOY shell model (Eqs. (3)-(4))
%for high value of
%the Prandtl number, $Pr = 10000$. Plotted is $< \theta_n >$ versus
%ln$k_n$ for the following values of teh constants: $N$ = 27, $\nu = 10^{-8}$,
%$D = 10^{-13, $f = \bar f = XXXX$. 
%\label{thdyn}}
%\end{figure}

\section{Scaling of eigenvalue spectra}
Now the stability of the Kolmogorov-Obukhov fixed point is examined 
in terms of the eigenvalue spectra.
The system eq.(\ref{goy}-\ref{passhell}) is written as
\begin{equation}
\left\{ \begin{array}{lll}
\dot{u} & = & f(u) \\
\dot{\theta} & = & g(u,\theta)
\end{array} \right.
\end{equation}

The Jacobian matrix is symbolically given by:
\begin{equation}
J =
\left ( \begin{array}{cc}
\frac{\partial f(u)}{\partial u} & 0 \\
\frac{\partial g(u,\theta)}{\partial u} & \frac{\partial
g(u,\theta)}{\partial \theta} \\
\end{array} \right)
\end{equation}
The matrix $\frac{\partial g(u,\theta)}{\partial u}$ will not
matter for the eigenvalues of $J$.
The eigenvalues of $\frac{\partial f(u)}{\partial u}$ were studied
in \cite{kada2} and \cite{Lohse}. It is most convenient to look 
at disturbances of phase
and modulus of the velocity variable $u_n = \bar{u}_n e^{i\phi_n}$: 
$\bar{u}_n =
u_n^{(0)} + \delta u_n$ and $\phi_n = \phi_n^{(0)} + \delta \phi_n$.
One then obtains for the stability of the modulus:
\begin{equation}
\dot{\delta u_n} = - \sum (D_{nm} + C_{nm}) \delta u_m
\end{equation}
The matrices $D_{nm}$ and $C_{nm}$ are the contributions of the dissipation
and cascade terms, respectively. Their expressions are:
\begin{equation}
D_{nm} = \nu k_n^2 \delta_{nm}
\end{equation}
and
\begin{eqnarray}
C_{nm} = \frac{\partial}{\partial u_m} (k_n u_{n+1} u_{n+2} - 
\delta k_{n-1} u_{n-1} u_{n+1} \nonumber \\ 
- (1-\delta) k_{n-2} u_{n-1} u_{n-2}).
\end{eqnarray}
where the index $(0)$ has been dropped, for convenience. 

The linearized equation for the variation of the phase is:
%\footnote{Here a similarity transformation has been used.}:
\begin{equation}
\dot{\delta \phi_n} = - \sum (D_{nm} - C_{nm}) \delta \phi_m
\end{equation}
 
\begin{figure}[htb]
\narrowtext
\unitlength= 0.0022\textwidth
\begin{center}
\begin{picture}(200,200)(0,0)
\put(0,0){\makebox(200,200){\epsfxsize=180\unitlength\epsffile{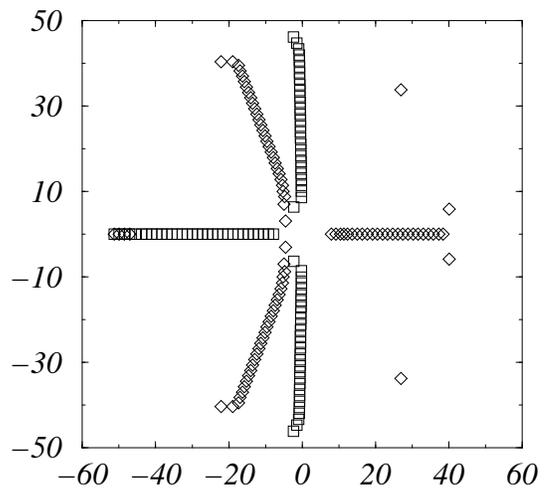}}}
\end{picture}
\end{center}
\caption{Eigenvalues of phase matrix ($\Diamond$) and amplitude matrix
($\Box$) for the velocity field in a polar plot for $\delta=0.5$.
\label{ueig}}
\end{figure}

Thus the stability eigenvalues of modulus and phases differ only in a
minus-sign in front of the cascade term.
We obtain similar results as in \cite{kada2}. For values of
$\delta < \delta_{bif} \approx 0.37$ all the eigenvalues of both phase
and modulus matrix have negative real part. Above this value, some of the real
eigenvalues of the phase matrix turn complex (in pairs) and cross the
imaginary axis. At $\delta=0.5$ they have become real again, but now
positive. This situation is shown in figure \ref{ueig}. The eigenvalues of the
modulus matrix eventually turn positive at $\delta \approx 0.7$.
In figure \ref{ueig} we have multiplied both real and imaginary part 
with a factor $p$ \cite{kada2}:
\begin{equation}
p=\frac{\log_2 (1+2^{10}|\sigma|)}{|\sigma|}
\end{equation}
Thus the phase remains unchanged, while the modulus is rescaled.
In case the eigenvalues have constant ratios, they are evenly spaced on the
plot. Thus we are able to visualize both the very small and very large
eigenvalues.

It is quite trivial to generalize this method to the passive scalar in
order to calculate the eigenvalues of $\frac{\partial
g(u,\theta)}{\partial \theta}$
and one finds: 
\begin{equation}
\dot{\delta \theta_n} = - \sum_m (D^{\theta}_{nm} + B_{nm}) \delta \theta_m
\label{thmod}
\end{equation}
where $D^{\theta}_{nm}$ is the dissipation term, very similar to the
one before
\begin{equation}
D^{\theta}_{nm} = D k_n^2 \delta_{nm}
\end{equation}
and $B_{nm}$ is the contribution
of the cascade term given by:
\begin{eqnarray}
B_{nm} = \frac12 \frac{\partial}{\partial \theta_m} ( k_n (u_{n-1}
\theta_{n+1} - u_{n+1}
\theta_{n-1}) \nonumber 
\\- k_{n-1} (u_{n-2} \theta_{n-1} + u_{n-1}
\theta_{n-2}) \nonumber \\
+ \frac12 k_{n+1} (u_{n+2} \theta_{n+1} + u_{n+1}
\theta_{n+2}) )
\end{eqnarray}
We see that $B_{nm}$ does not depend on $\theta$ as the $\theta$-variation
will be differentiated out.

For the phase disturbance $\delta \psi_n$
one gets
%\footnote{One performs again a
%similarity transformation.}:
\begin{equation}
\dot{\delta \psi_n} = - \sum_m (D^{\theta}_{nm} - B_{nm}) \delta
\psi_m
\label{thphas}
\end{equation}

We note that since $B_{n,n+2} = \frac12 k_{n+1} u_{n+1} = - B_{n+2,n}$ and
$B_{n,n+1} = \frac12 (k_{n+1} u_{n+2} + k_n u_{n-1}) = -B_{n+1,n}$, the matrix
$B_{nm}$ is antisymmetric and its eigenvalues will be purely
imaginary. The stability of phase and modulus will thus be the same.
This is actually a consequence of the conservation of $\sum |\theta
_n|^2$. In the model any cascade term $a_n u_{n+n'} \theta _{n+n''}$ has to be 
supplemented by a term $-a_{n-n''} u_{n+n'-n''} \theta_{n-n''}$, since
the conservation implies $\sum \theta_n \dot{\theta_n}=0$.
The first term gives a contribution to the matrix $B_{nm}$: 
$B_{n,n+n''}=a_n u_{n+n'}$, the second $B_{n,n-n''}=-a_{n-n''} u_{n+n'-n''}$
which corresponds to $B_{n+n'',n}=-a_n u_{n+n'}$. The matrix is thus
antisymmetric and this is not an artefact of the model since it stems from a 
conservation law also valid in a real system.

One does not expect to find any bifurcations since the diffusion 
matrix $-D^{\theta}_{nm}$ will cause the eigenvalues to have negative real
parts, whatever the value of the driving velocity. 

\begin{figure}[htb]
\narrowtext
\unitlength= 0.0022\textwidth
\begin{center}
\begin{picture}(200,200)(0,0)
\put(0,0){\makebox(200,200){\epsfxsize=180\unitlength\epsffile{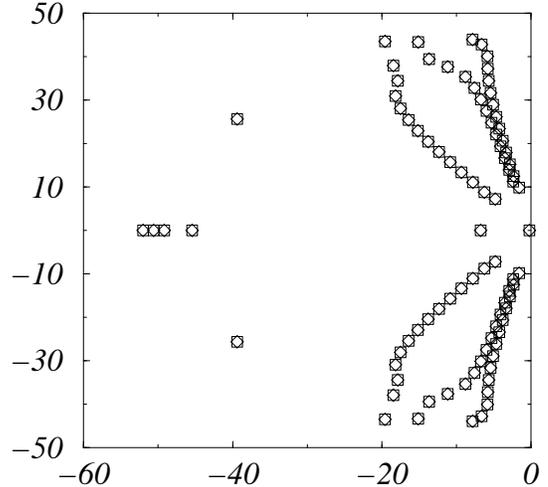}}}
\end{picture}
\end{center}
\caption{Eigenvalues of phase matrix ($\Diamond$) and amplitude matrix
($\Box$) for the passive field in a polar plot at $\delta=0.5$.
\label{theig}}
\end{figure}

Indeed one finds that for all values of $\delta$, the spectrum of eigenvalues of
the phase matrix 
is like shown in figure \ref{theig}, where the
eigenvalues are again multiplied by $p$. 
One observes evenly spaced eigenvalues organized in branches. The
presence of three branches (in both upper and lower half of the
complex plane) might be caused by the period three of $u_n$ (in $n$). 
If one inserts in the matrix $B_{nm}$ a fixed point of an imaginary
model without period 3, one just finds one branch in each half plane.

Let's us now consider what the above imply for the dynamics of the
model. At time $t=0$, starting from the particular state of the scalar
$\theta_n(0)$ we impose a small perturbation:
$\theta'_n(0)=\theta_n(0) + \delta \theta_n(0)$ and look how the
perturbation evolves in time, subjected to the {\it same} velocity field. 
For modulus and phase of $\delta \theta$ we have again equations
(\ref{thmod}) and (\ref{thphas}), since the matrices $D^{\theta}$ and
$B$ do not depend on $\theta$. 
Because the eigenvalues of $B-D^{\theta}$ have negative real
parts, the perturbation will damp out, meaning that after some time
$\theta'_n(t) \sim \theta_n(t)$ .
This has been observed already by Crisanti et al.\ \cite{cris93}.

\section{Intermittency}
The term intermittency is used in different contexts and a precise
mathematical definition does not exist. 
In turbulence one speaks of intermittency corrections to the
Kolmogorov or Obukhov-Corrsin power law.
In dynamical systems in general, intermittency means the presence
of quiescent periods randomly interrupted by burst.
It is believed that these phenomena are related: the scaling 
corrections should have their origin in intermittent behavior 
(in space and/or time) of velocity or energy dissipation.
Here we ask ourselves the question whether the more pronounced deviations 
from classical scaling for the passive scalar are reflected by a more
intermittent behavior (in time) of the passive scalar field.
In order to test this we invoke a definition of intermittency for dynamical 
systems proposed by Benzi et al.\ \cite{bppv85}.  
 
Firstly, the response function $R_t(\tau)$ is defined as the rate at which
a disturbance  vector $\delta \vec{x}(t)$ of the system 
$\dot{\vec{x}}=f(\vec{x})$ 
has expanded after a time $\tau$.
\begin{equation} R_t(\tau) = \frac{|\delta \vec{x}(t+\tau)|}{|\delta
\vec{x}(t)|}
\end{equation}
In chaotic systems one typically observes that $~~~~~~~~$ 
$|\delta \vec{x}(t+\tau)| \sim |\delta \vec{x}(t)| e^{\tau \lambda}$, 
where $\lambda$ is the maximum Lyapunov exponent. 
This maximum Lyaponov exponent can be calculated by averaging the 
{\em instantaneous} Lyapunov exponents $\frac{\ln R_t(\tau)}{\tau}$:
$\lambda = \lim_{\tau \to \infty} \frac{\langle \ln R(\tau) \rangle}{\tau}$. 

Intermittency can be thought of as connected to the fluctuations
of the instantaneous Lyapunov exponent. An intermittency parameter $\mu$ is thus 
defined as the {\em variance} of $\ln R(\tau)$ \cite{bppv85}:
%
%Generalized Lyapunov exponents $L(q)$ are related to the moments of
%the averaged response function:
%\begin{equation}
%\langle R(\tau)^q \rangle \sim \exp(L(q)\tau)
%\end{equation}
% 
%It can be shown for the maximum Lyapunov exponent $\lambda =
%\frac{\langle
%\ln R(\tau) \rangle}{\tau}$:
%\begin{equation}
%\lambda = \frac{d L(q)}{d q}|_{q=0}
%\end{equation}
%	
%In absence of fluctuations of $R(\tau)$ one has:
%\begin{equation}
%L(q)=\lambda q
%\end{equation}
%
%Benzi et al. argue that intermittency is connected to the fluctuations
%of the local Lyapunov exponent. It will be almost zero during the
%quiescent periods, but much larger when a burst occurs.
%Then the intermittency parameter $\mu$ is defined as the variance of $\ln
%R(\tau)$:
\begin{equation}
\langle (\ln R(\tau))^2 \rangle  - \langle \ln R(\tau) \rangle^2 = \mu
\tau
\end{equation}
% 
%This is equivalent to:
%\begin{equation}
%\mu = \frac{d^2 L(q)}{d q^2}|_{q=0}
%\end{equation}
%
%These quantities are calculated using standard techniques \cite{ben}.
%The equations of motion (\ref{goy}-\ref{passhell}) are numerically integrated 
%along the equations for the tangent vector $\dot{\delta x}=J(x) \delta
%x$ where $J$ is the Jacobian matrix. The tangent vector initially has
%unit length. In order to avoid numerical overflow we renormalize the
%vector at times $t=j \tau$ and note its length $\alpha_j$. Then the
%Lyapunov exponent is given by:
%\begin{equation} \label{lam1}
%\lambda = \lim_{l \rightarrow \infty} \frac{1}{l\tau} \sum_{j=1}^l
%\ln \alpha_j
%\end{equation}
%The expression for the intermittency parameter $\mu$ is: 
%\begin{equation} \label{calcmu}
%\mu = \varsigma - \lambda^2 \tau
%\end{equation}
%where $\varsigma$ is given by:
%\begin{equation}
%\varsigma =\frac{\langle (\ln R(\tau))^2 \rangle}{\tau}
%= \lim_{l \rightarrow \infty}
%\frac{1}{l\tau} \sum_{j=1}^l (\ln \alpha_j)^2
%\end{equation}

Numerically we proceed as follows: the equations of motion 
(\ref{goy}-\ref{passhell}) are numerically integrated 
along the equations for the disturbance vector $\dot{\delta {\vec{x}}}=
J({\vec{x}}) \delta
{\vec{x}}$ where $({\vec{x}}) = (u,\theta)$ and $J$ is the Jacobian matrix.
We then estimate the response function and the first two 
cumulants of its logarithm.
From time to time the disturbance vector is renormalized in order to avoid 
numerical overflow \cite{ben}.

Consider now the Lyapunov exponents of the full system of
velocity and passive scalar field (\ref{goy}-\ref{passhell}), 
with the parameters (\ref{standpar}).
We distinguish how the perturbation, initially
applied equally on {\it both} systems ($\delta u_n =\delta \theta_n = constant$
for all $n$), evolves independently in each of the two parts of the system.
Therefore we introduce response functions that measure the expansions
on the velocity part $R^{(u)}_t(\tau)$ and on the scalar part
$R^{(\theta)}_t(\tau)$.
They are defined as:
\begin{eqnarray}
R^{(u)}_t(\tau) = \frac{|\delta u(t+\tau)|}{|\delta u(t)|} \nonumber
\\
R^{(\theta)}_t(\tau) = \frac{|\delta \theta(t+\tau)|}{|\delta
\theta(t)|}
\end{eqnarray}
The corresponding maximal Lyapunov exponents is defined as:
\begin{equation}
\lambda^{(u)} = \frac{\langle \ln R^{(u)}(\tau) \rangle}{\tau}
\end{equation}
and the analogue for $\lambda^{(\theta)}$.
This is reminiscent to the ``Eulerian Lyapunov exponent'' and
``Lagrangian Lyapunov exponent'' introduced by Crisanti et
al.\cite{cris91,cris93}. However the Lagrangian behavior of the particles 
is only equivalent to the Eulerian passive scalar field in the case of 
vanishing diffusion. 

The intermittency parameters are related to the variance of $\ln
R(\tau)$ is the same way as before.
We find numerically:
\begin{equation}
\lambda^{(u + \theta)} \approx \lambda^{(u)} \approx
\lambda^{(\theta)} = 0.165 \pm 0.002 
\end{equation}
where we have denoted the Lyapunov exponent for the full system as:
$\lambda^{(u + \theta)}$.
On expects theoretically:
\begin{equation}
\lambda^{(u + \theta)} = \max [\lambda^{(u)},
\lambda^{(\theta)}]
\end{equation}
since the disturbance vector will evolve towards the most expanding
direction. Our result is obviously in agreement with this relation.
For the Eulerian and Lagrangian Lyapunov exponent, one has numerically 
found the generic inequality $\lambda_L \ge \lambda_E$. 
Our result $\lambda^{(u)} \approx \lambda^{(\theta)}$ might be due to the
presence of a finite value of diffusivity.  

The intermittency parameters $\mu$ differ on the other
hand significantly from each other.
We find (for $\delta=0.5$): $\mu^{(\theta + u)} =  0.127 \pm 0.003$,
$\mu^{(u)} = 0.125 \pm 0.003$ and $\mu^{(\theta)} = 0.151 \pm 0.003$.
Thus the passive scalar behaves equally chaotic, but 
more intermittent, than the velocity.
In order to understand this we write the equation for $\theta$ as
follows:
\begin{equation}
\dot{\theta}=A(u) \theta + f
\end{equation}
where A is a matrix depending on $u$, whose eigenvalues $\sigma$ fulfil
the inequality $\Re \sigma < 0$. Thus, if $u$ is constant in time,
$\theta$ will converge to its fixed point $\theta = - A(u)^{-1} f$. 
Alternatively, if $u$ fluctuates as for $\delta=0.5$ in the GOY model,
this ``fixed point'' will fluctuate as well. Moreover the convergence 
towards it will be irregular since the eigenvalues of $A(u)$ vary. 
It is therefore natural to expect that the behavior of $\theta$ is more
intermittent than that of $u$. 

%Equation .. has formal solution:
%\begin{equation}
%\theta= e^{\int A(t) dt} \int e^{-\int A(t) dt} dt f + e^{\int A(t) dt} \theta _0
%\end{equation} 

%Crisanti et al. \cite{cris93} found, measuring predictability times
%(i.e. the time at which a perturbation exceeds a certain threshold
%value) that the Lagrangian Lyapunov exponent is larger than the
%Eulerian one. This must somehow be due to the difference in definition of
%the Lyapunov exponent but we do not know how exactly.
%We don't know the origin of this discrepancy.

\section{Conclusions and outlook}
We studied various properties of a shell model for the
advection of a passive scalar. We calculate the eigenvalue spectrum of
the Kolmogorov fixed point and show that the passive scalar part is
stable against perturbations. This is in fact not very surprising
since it follows from conservation of $\sum |\theta|^2$.

The fixed point does on the other hand not follow the Obukhov-Corrsin scaling. 
This suggests that the scaling 
properties of the passive scalar are much more sensitive, than the velocity, 
to non-inertial properties such as the dissipation and perhaps
also the forcing. 
In experiments the scaling zones of the passive scalar are less
clearly observed and some
``nonuniversality'' (with respect to forcing)
seems to have been observed in \cite{pinton}, but 
we of course cannot pretend that the mechanism observed here will be found also
in real systems.

We have measured the relation between Lyapunov exponents and 
intermittency parameters of the passive scalar and velocity. We find:
$\lambda^{(u)} \approx \lambda^{(\theta)}$ and $\mu^{(u)}< \mu^{(\theta)}$.
Thus the passive scalar behaves more intermittently which is in agreement 
with its more pronounced deviations from linear scaling. 
It would be interesting to establish these relations from the 
equations of motion. 

It could also be interesting to investigate the relation between coherent 
structures in the GOY model \cite{Fridolin,Gilson} and those in the passive
scalar model.

\section{Acknowledgments}
J.K. is grateful to the Niels Bohr Institute for warm hospitality.
Discussions with Hugues Chat\'e, Ted Janssen, Ana\"el Lema\^\i tre and 
Paolo Muratore-Ginanneschi are kindly acknowledged. We thank Detlef Lohse
for providing us with the code to extract the Kolmogorov fixed
point.
% --------------------------------------------------------------------
% BIBLIOGRAPHY
% --------------------------------------------------------------------

% --------------------------------------------------------------------

\end{multicols}
\end{document}